\documentclass[11pt]{article}
\usepackage{graphicx}

\newcommand{\BABARPubYear}    {03}

\newcommand{\BABARConfNumber} {16}
\newcommand{\SLACPubNumber} {10105}
\newcommand{\LANLNumber} {0308027}

\input babarsym 
\newcommand{\bi}{\begin{itemize}}
\newcommand{\ei}{\end{itemize}}
\newcommand{\ben}{\begin{enumerate}}
\newcommand{\een}{\end{enumerate}}
\newcommand{\bc}{\begin{center}}
\newcommand{\ec}{\end{center}}
\newcommand{\bt}{\begin{table}}
\newcommand{\et}{\end{table}}
\newcommand{\be}{\begin{equation}}
\newcommand{\eeq}{\end{equation}}

\newcommand{\ba}{\begin{eqnarray}}
\newcommand{\ea}{\end{eqnarray}}

\newcommand{\vs}{\vspace}
\newcommand{\la}{\ifmmode {\leftarrow} \else {$\leftarrow$}\fi}
\newcommand{\Ra}{\ifmmode {\Rightarrow} \else {$\Rightarrow$}\fi}
\newcommand{\La}{\ifmmode {\Leftarrow} \else {$\Leftarrow$}\fi}
\newcommand{\Lra}{\ifmmode {\Longrightarrow} \else {$\Longrightarrow$}\fi}
\newcommand{\Lla}{\ifmmode {\Longleftarrow} \else {$\Longleftarrow$\fi}}
\newcommand{\Llra}{\ifmmode {\Longleftrightarrow} \else {$\Longleftrightarrow$\fi}}
\newcommand{\Lk}{\ifmmode {{\cal L}} \else {${\cal L}$}\fi}
\newcommand{\Wt}{\ifmmode {{\cal W}} \else {${\cal W}$}\fi}
\newcommand{\Br}{\ifmmode {{\cal B}} \else {${\cal B}$}\fi}
\newcommand{\N}{\ifmmode {{\cal N}} \else {${\cal N}$}\fi}
\newcommand{\G}{\ifmmode {{\cal G}} \else {${\cal G}$}\fi}
\newcommand{\E}{\ifmmode {{\cal E}} \else {${\cal E}$}\fi}
\newcommand{\Pfr}{\ifmmode {{\cal F}} \else {${\cal F}$}\fi}
\newcommand{\Aone}{\ifmmode {{\cal A}_1} \else {${\cal A}_1$}\fi}
\newcommand{\rha}{\ifmmode{\mbox{\rho^2_{{\cal A}_1}}} \else {\mbox{$\rho^2_{{\cal A}_1}$}}\fi}
\newcommand{\rhf}{\ifmmode{\rho^2_{\cal F}}\else{\mbox{$\rho^2_{\cal F}$}}\fi}
\newcommand{\om}{\ifmmode {w} \else {$w$}\fi}
\newcommand{\dom}{\ifmmode {\Delta w} \else {$\Delta w$}\fi}
\newcommand{\tBz}{\ifmmode {\tau_{\Bz}} \else {$\tau_{\Bz}$}\fi}
\newcommand{\tBp}{\ifmmode {\tau_{\Bu}} \else {$\tau_{\Bu}$}\fi}
\newcommand{\BtoDs}{\mbox{$\bar{B^0}\rightarrow D^{*+} \ell^- \bar{\nu_\ell}$}}
\newcommand{\BtoDss}{\mbox{$\bar{B}\ra\Dstarp\ell^-\bar{\nu_\ell}\X$}}

\newcommand{\pstar}{\ifmmode {\pi_{\rm slow}} \else {$\pi_{\rm slow}$}\fi}
\newcommand{\pstarp}{\ifmmode {\pi^+_{\rm slow}} \else {$\pi^+_{\rm slow}$}\fi}
\newcommand{\plab}{\ifmmode{p} \else {$p$} \fi}
\newcommand{\ctdl}{\ifmmode{ \cos(\theta_{BY}) } \else {$\cos(\theta_{BY})$} \fi}
\newcommand{\ks}{\ifmmode{k^*} \else {$k^*$} \fi}
\newcommand{\mnusq}{\ifmmode{{M_\nu}^2} \else {$M_{\nu}^2$}\fi} 
\newcommand{\DTau}{\ifmmode {\Delta \tau} \else {$\Delta \tau$}\fi}
\newcommand{\ggcc}{\ifmmode {GeV^2/c^4} \else {$GeV^2/c^4$}\fi}
\def\BpBm {\ensuremath{B^+ {\kern -0.16em \Bub}}}

\newcommand{\bkg}{{background}\xspace}
\newcommand{\Ddstar}{\ensuremath{D^{**}}\xspace}

\newcommand{\dfrac}[2]{\frac{\displaystyle #1}{\displaystyle #2}}

\newcommand{\AoneVcb}{\ifmmode {{\cal A}_1(1)|V_{cb}|}\else {${\cal A}_1(1)|V_{cb}|$}\fi}
\newcommand{\omt} {\ifmmode {\tilde{w}} \else {$\tilde{w}$} \fi}

\newcommand{\Dsp}{\ifmmode{D^{*+}} \else {$D^{*+}$} \fi}
\newcommand{\dm}{\ifmmode{\Delta M} \else {$\Delta M$} \fi}
\newcommand{\D} {\ifmmode{\Delta} \else {$\Delta$} \fi}
\newcommand{\TBY}{\ifmmode{\theta_{\Bz, D^*\ell}} \else {$\theta_{\Bz, D^*\ell}$} \fi}
\newcommand{\Kpi}{\ifmmode{K^-\pi^+} \else {$K^-\pi^+$} \fi}
\newcommand{\Kppp}{\ifmmode{K^-\pi^+\pi^+\pi^-} \else {$K^-\pi^+\pi^+\pi^-$} \fi}
\newcommand{\Kppz}{\ifmmode{K^-\pi^+\pi^0} \else {$K^-\pi^+\pi^0$} \fi}
\newcommand{\Kspp}{\ifmmode{K_S\pi^+\pi^-} \else {$K_S\pi^+\pi^-$} \fi}
\long\def\inst#1{\par\nobreak\kern 4pt\nobreak
    {\it #1}\par\vskip 10pt plus 3pt minus 3pt}

\begin{document}
{\pagestyle{empty}

\begin{flushright}
\babar-CONF-\BABARPubYear/\BABARConfNumber \\
SLAC-PUB-\SLACPubNumber \\
hep-ex/\LANLNumber \\
August 2003 \\
\end{flushright}

\par\vskip 3cm

% Title of the paper
\begin{center}
\Large \bf Measurement of \boldmath{\Vcb} 
using $\bar{B^0} \rightarrow D^{*+} \ell^- \bar{\nu_\ell}$ Decays
\end{center}
\bigskip

\begin{center}
\large The \babar\ Collaboration\\
\mbox{ }\\
\today
\end{center}
\bigskip \bigskip

% Abstract
\begin{center}
\large \bf Abstract
\end{center}
A preliminary measurement of \Vcb and the branching fraction 
\mbox{${\cal B}(\bar{B^0} \rightarrow D^{*+} \ell^- \bar{\nu_\ell}$)} 
has been performed based on a sample of about 55,700 
\mbox{$\bar{B^0} \rightarrow D^{*+} \ell^- \bar{\nu_\ell}$} decays 
recorded with 
the \babar\  detector. 
The decays are identified in the 
\mbox{$D^{*+} \rightarrow D^0 \pi^+$} final state, with the
$D^0$ reconstructed in three different decay modes.  The differential
decay rate is measured as a function of the relativistic boost of the
$D^{*+}$ in the ${\bar{B^0}}$ rest frame.  The value of the differential
decay rate at `zero recoil', namely 
the point at which the $D^{*+}$ is at rest in the ${\bar{B^0}}$
frame, is predicted in Heavy Quark Effective Theory 
as a kinematic factor times ${\cal F}(1)|V_{cb}|$, where ${\cal F}$ is
the unique form factor governing the decay.  We 
extrapolate the measured differential decay rate to the zero recoil point and
obtain ${\cal F}(1)|V_{cb}|=(34.03\pm0.24\pm1.31)\times
10^{-3}$.  Using a theoretical calculation for ${\cal F}(1)$ we extract
$$
\nonumber |V_{cb}| = (37.27\pm0.26 \mathrm{(stat.)} \pm1.43 \mathrm{(syst.)} ^{+1.5} _{-1.2} \mathrm{(theo.)})\times 10^{-3} .
$$
From the integrated decay rate we obtain 
$$
\nonumber {\cal B}(\bar{B^0} \rightarrow D^{*+} \ell^- \bar{\nu_\ell})=(4.68 \pm 0.03 \pm 0.29)\%.
$$

\vfill
\begin{center}
Contributed to the 
XXI$^{\rm st}$ International Symposium on Lepton and Photon Interactions at High~Energies, 8/11 --- 8/16/2003, Fermilab, Illinois USA
\end{center}

\vfill
\bigskip \bigskip \bigskip
\begin{center}
{\em This work is dedicated to the memory of Paolo Poropat.}
\end{center}
\bigskip \bigskip

\vspace{0.4cm}
\begin{center}
{\em Stanford Linear Accelerator Center, Stanford University, 
Stanford, CA 94309} \\ \vspace{0.1cm}\hrule\vspace{0.1cm}
Work supported in part by Department of Energy contract DE-AC03-76SF00515.
\end{center}

\newpage
} % end of pagestyle{empty}

% Input author list file
% \input authors_eps2003.tex
\begin{center}
\small

The \babar\ Collaboration,
\bigskip

%% author list as of 02-Jun-2003 (595 authors)
%
B.~Aubert,
R.~Barate,
D.~Boutigny,
J.-M.~Gaillard,
A.~Hicheur,
Y.~Karyotakis,
J.~P.~Lees,
P.~Robbe,
V.~Tisserand,
A.~Zghiche
\inst{Laboratoire de Physique des Particules, F-74941 Annecy-le-Vieux, France }
% ...

\end{center}\newpage

% The body of the paper starts here

%\tableofcontents

\clearpage

\setcounter{footnote}{0}

\section{Introduction}

In the framework of the Standard Model, the weak coupling between
quarks of different flavors is described by the
Cabibbo-Kobayashi-Maskawa (CKM) matrix. %\cite{CKM} matrix.  
%The elements of this unitary matrix are not predicted by theory, apart
%from the constraints from the unitarity of the matrix.\par
Its elements are not predicted by theory, but are constrained  only to
give a unitary matrix.

A precise measurement of \Vcb, the element corresponding to the $b\ra
c$ transitions, is needed to determine whether the CKM
matrix provides a quantitatively accurate description of
%constrain the parameters which describe the process of 
the \CP\ violation observed in \Bz mesons.  Progress in the
phenomenological description of heavy flavor semileptonic decay allows
the determination of \Vcb with small theoretical uncertainty, either
from the inclusive process $b \rightarrow c\ell \bar\nu$, or from an
analysis of the form factors in the decay \BtoDs. The present
measurement is based on the second approach~\cite{method}.\par 

The decay rate for the process is proportional to $\Vcb^2$ and to the
square of the hadronic matrix elements describing the transition from a
\Bzb to a \Dsp meson.  In the context of the Heavy Quark Effective
Theory (HQET)~\cite{F1}, the matrix element is proportional to a
single form factor \Pfr(\om), where \om\ is the product of the \Bzb
and \Dsp four-vector velocities. For $\om=1$, the \Dsp is produced at
rest in the \Bzb rest frame.  Heavy-flavor symmetry predicts the
normalization $\Pfr(1) = 1$ in the limit of an infinitely massive $b$-quark.  
Corrections to this prediction due to
perturbative QCD have been computed up to second order in
\as~\cite{QCDII} and the effect of finite $b$ and $c$ quark masses has
been calculated in the framework of HQET, yielding $\Pfr(1) =
0.913^{+0.030}_{-0.035}$~\cite{Hashimoto}.\par 

A measurement of the differential decay rate near $\om=1$ determines
\Vcb with small theoretical uncertainty. 
%Due to phase space
%suppression, this quantity is determined by extrapolating to $\om=1$
%the differential rate d$\Gamma$/d\om, where \Pfr(\om) is parametrized
The rate at $\om=1$ is suppressed by phase space.  Consequently, the
differential rate d$\Gamma$/d\om\ is measured, where \Pfr(\om) is
parametrized using several different functional
forms~\cite{Neuold,Neunew} (see also discussion below), and is extrapolated
to $\om=1$.  Results based
on this approach have been reported by the ARGUS~\cite{ARGUS},
Belle~\cite{BELLE}, and CLEO~\cite{CLEO} collaborations operating at
the $\Upsilon$(4S) and by ALEPH~\cite{ALEPH1,ALEPH2},
DELPHI~\cite{DELPHI1,DELPHI:vcb} and OPAL~\cite{OPAL} at LEP.\par 

This paper presents preliminary results for the measurements of
\Vcb and of the branching fraction ${\cal B}($\BtoDs) performed on a
sample of \BtoDs\ decays selected from events with a high momentum
charged lepton $\ell^-$ and a $\Dsp$~\footnote{Charge conjugate states
are always implicitly considered. Lepton as used here means either
electron or muon.}.  The \Dsp is reconstructed from its decay to a
charged pion and a \Dz. This pion is produced with small momentum, and
is commonly referred to as the slow pion (\pstar).

This paper is organized as follows. The next section is dedicated to a
brief description of the \babar\ detector and of the data sets
employed. Section~\ref{sec:selection} describes the event selection
and the composition of the sample used for the measurement.  The fit
method and its results are described in Section~\ref{sec:fit}; the
evaluation of the systematic error is discussed in
Section~\ref{cha:sys}. Finally, Section ~\ref{conc} presents the conclusions
and the comparison with other measurements of \Vcb.

\section{The \babar\ Detector }

The data were collected with the \babar\ detector at the \pep2\
asymmetric energy \epem storage ring~\cite{ref:pep2}.
%The Lorentz boost $\beta\gamma = 0.55$ at the \FourS.
The sample consists of 79.1~\invfb collected at a center-of-mass
energy corresponding to the \FourS resonance (``on-resonance''), and
9.6~\invfb collected 40~\mev below the \FourS resonance
(``off-resonance'') for continuum background studies. The on-resonance
sample corresponds to $N_\Upsilon = (85.9\pm 0.9) \times 10^6$
\FourS decays to \BB mesons.\par
%The \FourS is boosted along the collision axis with a Lorentz factor $\beta\gamma = 0.55$.\\
The \babar\ detector is described in detail elsewhere~\cite{ref:babar}.
The momenta of charged particles are measured using a combination of 
a five-layer silicon vertex tracker (SVT) and a 40-layer drift chamber (DCH)
in a 1.5~T solenoidal magnetic field.
A detector of internally-reflected Cherenkov radiation (DIRC) is used
for charged-particle identification.
Kaons are identified with a neural network based on the likelihood ratios 
calculated from d$E$/d$x$ measurements in the SVT and DCH, and from the 
information from the DIRC.
A finely segmented CsI(Tl) electromagnetic calorimeter (EMC) is used to detect 
photons and neutral hadrons, and to identify electrons.
Electron candidates are required to have a ratio of EMC energy to track 
momentum, EMC cluster shape, DCH d$E$/d$x$, and DIRC Cherenkov angle all 
consistent with the electron hypothesis. 
The instrumented flux return (IFR) contains resistive plate chambers for muon 
and long-lived neutral hadron identification.
Muon candidates are required to have IFR hits located along the extrapolated 
DCH track, and energy deposition in the EMC and penetration length in the IFR
consistent with a minimum ionizing particle.

\section{Event Selection and Sample Composition}
\label{sec:selection}

\subsection{Event Selection}

Events are selected that contain a fully-reconstructed \Dsp and an oppositely
charged lepton. 
%The \BtoDs~ signal is enhanced by means of kinematic cuts.
Those $\Dsp\ell^-$ combinations not coming from the signal decay are
suppressed by means of kinematic cuts.
%Background control samples are also selected.

A candidate lepton is searched for among all charged tracks with
momentum greater than 1.2~GeV/$c$ in the \FourS rest frame. Electron
candidates are selected with an efficiency of about 90\%
and a hadron misidentification probability of less than
0.2\%.  
Muon candidates are selected with an efficiency of $\simeq 60$\%
%The efficiency for selecting a muon decreases over the years
%from 70\% to 50\% 
and hadron misidentification probability is about 2\%.

The \Dsp candidate is selected in the decay mode $\Dsp\ra\Dz\pi^+$ and
the \Dz meson is reconstructed in the three modes $\Kpi$, $\Kppp$, and
$\Kppz$. The $\piz$ is reconstructed from two photons, each with energy
larger than 30~\mev, and must have a total energy larger than 200~\mev and an
invariant mass between 119.2 and 150.0~\mevcc.  The invariant mass of
the two photons is constrained to the $\piz$ mass and the pair is kept
as a $\piz$ candidate if the \chisq probability of the fit exceeds
1\%.  Charged kaon candidates satisfy loose identification criteria
for the $\Kpi$ mode and tighter criteria for the $\Kppp$ and $\Kppz$
modes.  For the $\Kppz$ mode, the \Dz candidate is retained if the
square of the decay amplitude in the Dalitz plot for the three-body
candidate, based on measured amplitudes and phases~\cite{ref:E687}, is
larger than 10\% of its maximum value. %across the Dalitz plot.
\Dz candidates are accepted if they have an invariant mass within 17~\mevcc of the \Dz mass
for the $\Kpi$ and $\Kppp$ modes, and within 34~\mevcc for the 
$\Kppz$ mode.
The invariant mass of the decay products is then constrained to the \Dz mass ~\cite{PDG} 
and the tracks are constrained to a common vertex by a simultaneous fit.  
The \Dz candidate is retained if the \chisq probability of the fit
is more than 0.1\%.
The low-momentum pion candidates for the $\Dsp\ra\Dz\pstarp$ 
decay are selected from among tracks with total momentum in the laboratory frame less than 
450 \mevc, and momentum transverse to the beam line in the laboratory frame 
greater than 50 \mevc.
The momentum of the \Dsp candidate in the \FourS rest frame 
is required to be between 0.5 and 2.5 \gevc. These requirements retain
essentially all signal events and reject higher momentum \Dsp from 
continuum events.
Candidates from \Dsp  are %retained in the final selection 
%if the mass difference \mbox{$\Delta M  = M_{\Dz\pstar}-M_{\Dz}$} is less 
%than 165~\mevcc, where $M_{\Dz\pstar}$ is calculated after constraining the \Dz mass.
preselected with the cut on the mass difference \mbox{$\Delta M  = M_{\Dz\pstar}-M_{\Dz}< 165 ~\mevcc$}.
The $\Delta M$ distribution has a kinematic
threshold at the mass of the $\pi^+$, and a peak at 145.5 \mevcc
with a resolution of about 1 \mevcc .
% We use events in the sideband of the \deltam distribution in order to evaluate
% the combinatoric background below the signal. 
The combinatoric background below the signal is evaluated using events in the
\deltam\ sideband.
A harder \dm cut is applied later. 

The \chisq probability of the fit to 
a common vertex of the lepton, the \pstar, and the \Dz candidate, 
constrained to the beam spot, must be greater than 1\%.

Candidates for $\Dsp\ell^-$  are selected if 
$|\cos\Delta\theta^*_{\rm thrust}| < 0.85$, where $\Delta\theta^*_{\rm thrust}$ 
is the angle between the thrust axis of the $\Dsp\ell^-$ candidate and the 
thrust axis of the remaining charged and neutral particles in the event. 
The distribution of $|\cos\Delta\theta^*_{\rm thrust}|$ is peaked at 1 for 
jet-like continuum events, and is flat for $B\overline B$ events, where 
the two $B$ mesons decay independently and practically at rest with 
respect to each other. This cut retains about 85\% of signal and rejects about 47\% of continuum
events. 

Finally, the angle between the lepton and the \Dsp in the \FourS rest frame must %Bob
satisfy $\cos\theta_{D^*\ell}<0$. This cut reduces significantly the
events due to random combinations of a lepton and a \dsp produced by
two different B mesons, while keeping most of the signal events.
%With all the above requirements satisfied, a sample of 55,700 events is selected in the signal
%region.
\subsection{Sample Composition} 

Several processes contribute to the selected sample:

\begin{enumerate}
\item Signal decays.

\item Combinatoric background from any random combination of tracks mimicking
a true \Dz or from a correctly identified \Dz,  combined with a
low-momentum charged track not originating from \Dsp decay.  This
category includes events from $B\bar{B}$ and from continuum
production.  Because of its combinatorial origin, the $\Delta M$
distribution does not exhibit a peak.
% due to true $\Dsp\pi$ combinations.
 All the other backgrounds listed below peak in \dm.

\item Fake-lepton background, in which a charged hadron is misidentified 
as a lepton and is combined with a correctly reconstructed $\Dsp$. Due
to the very good performance of the electron identification, the rate of
fake electrons is negligible.

\item Continuum background, where a true \Dsp\ and a lepton (true or fake) are produced from $\epem\to c\bar{c}$ events.

\item ``Uncorrelated'' background, in which the lepton and the \Dsp originate from two different $B$ mesons.  
%This background is considerably
%reduced by the requirement that the \Dsp and the lepton have opposite
%charge. The residual background is due to several processes: events in
Most of these combinations are removed by the requirement that the
lepton and \Dsp have opposite charge.  Opposite charge combinations
arise mainly from events with \BzBzb oscillations where both B mesons
decay into the same flavor eigenstate, or events in which the \Dsp
comes from the hadronization of the virtual $W$ from $B$ decay, and thus has a charge
opposite to the one expected from its parent B.

\item \BtoDss~ background, where one (or more) hadrons are produced in addition to the $\Dsp$, 
for either neutral or charged $B$. 
A considerable fraction of this background is due to the intermediate production 
of orbitally excited $D$ states, which then decay to a $\Dsp\pi$ final state; this 
is referred to as \Ddstar background in the following.

\item ``Correlated'' background, in which the $B$ decays to a \Dsp and a heavy particle 
(either a charm meson or a $\tau$), which then decays to a lepton. 
These events have the same charge correlation as signal events, 
but are suppressed by their low branching fraction and by the kinematic cuts.
Such events correspond to a few percent of the peaking sample. Their contribution is fixed 
to the value computed in the simulation.

\end{enumerate}

\begin{figure}[!ht]
\begin{center}
\includegraphics[width=3.2in]{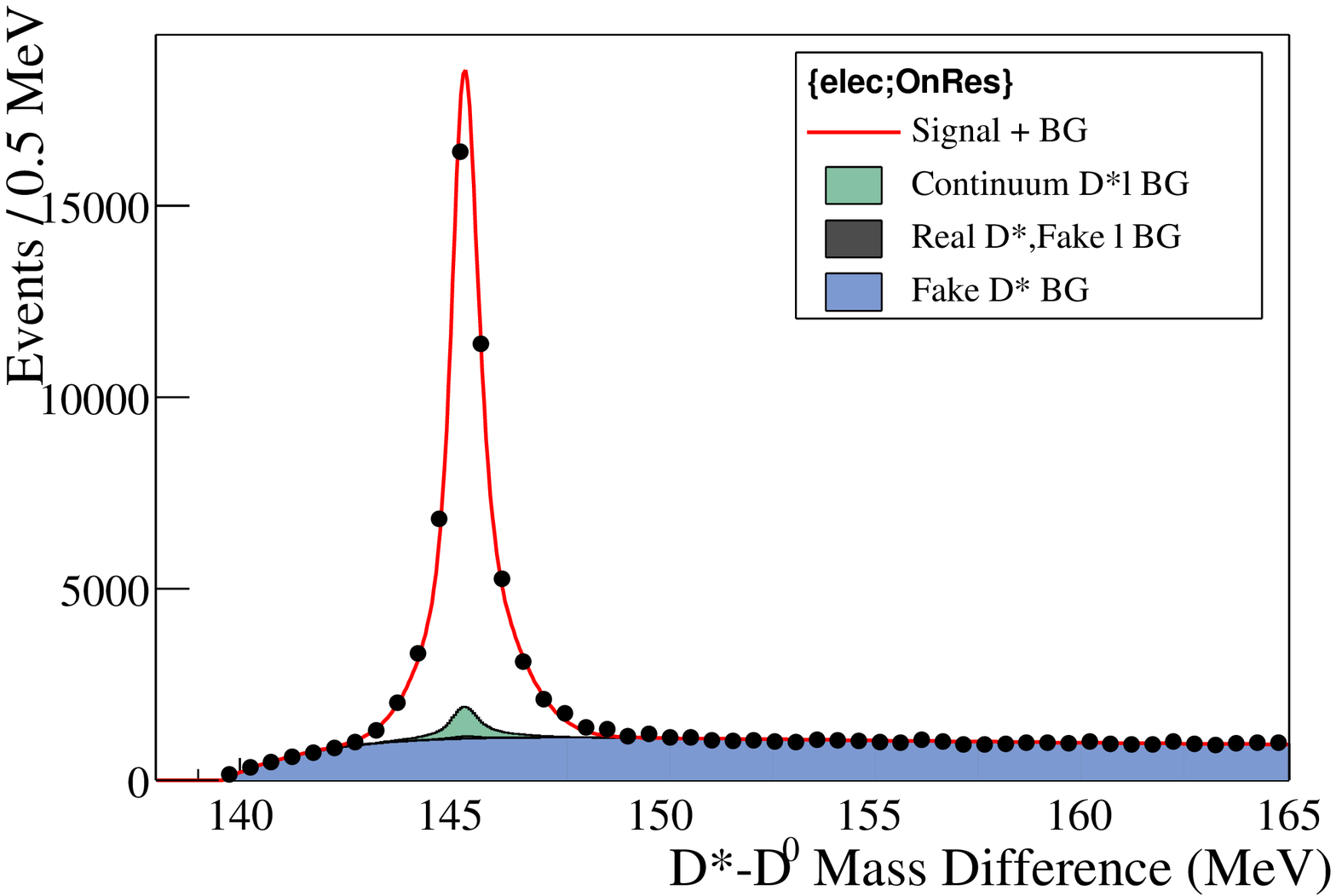}
\includegraphics[width=3.2in]{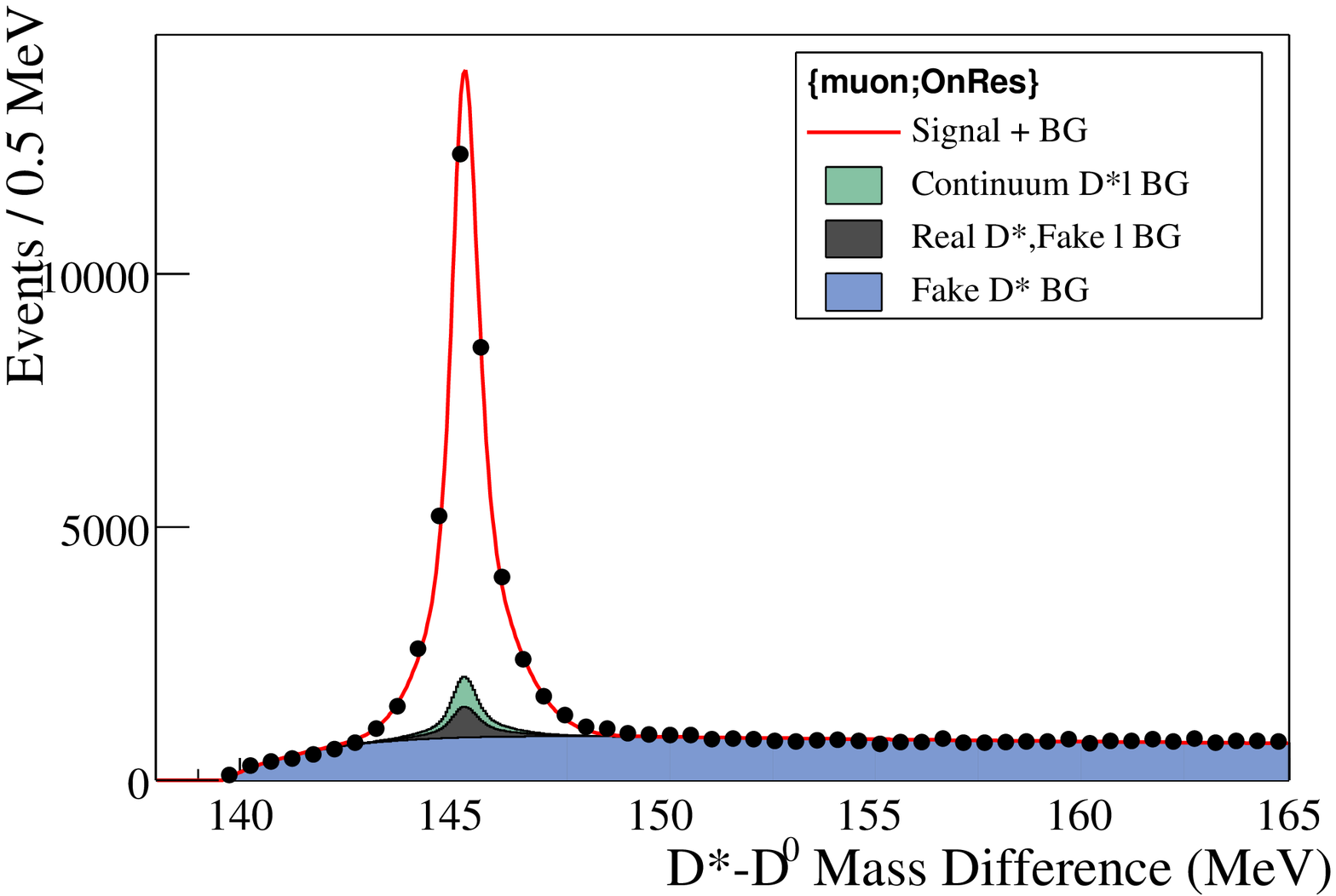}\\
\caption{\deltam distribution for events passing all selection criteria: 
$\Dstarp e^-$ (left) and $\Dstarp\mu^-$ (right) candidates in {\em on-resonance} data. 
Points correspond to data, and the curve is the result of unbinned 
maximum likelihood fits to this signal sample, the fake-lepton control sample and the
rescaled off-resonance events. The shaded distributions correspond to the {\em continuum,  
fake-lepton}, and {\em combinatorial} backgrounds described in the text. 
The {\em correlated-lepton}, the \mbox{$\bar{B}\ra\Dstarp\ell^-\bar{\nu_\ell}X$} and the 
{\em uncorrelated} backgrounds are included in the signal distribution.}
\label{f:dm}
\end{center}
\end{figure}

%The measurement is performed in several stages. First, the \dm
%distribution is fit to the sum of mostly signal events in the peak and
%a broad background distribution.  The amount of fake leptons is
%computed from a data control sample and the amount of events from
%continuum is computed by fitting the rescaled off-resonance
%events. Then only the events in a restricted \dm range are further
%considered. The fractions of uncorrelated and
%\Ddstar background in the selected set are determined from a fit to $ \cos\TBY $ (see below for its definition),
%while the residual correlated background is computed from the simulation. 
%These fits are  performed separately for ten different bins of \omt 
%(\omt\ is an estimator of \om\ defined below).
%It should be noted that while the integrated amount 
%of continuum background events is computed  from off-resonance data,
%their \omt\ shape is obtained from the simulation, 
%due to the small size of the off-resonance data sample. 
%The resulting signal distribution is then fitted to
%extract \AoneVcb\ and \rha . A detailed description of this procedure is reported in the
%following.  

%The composition of the sample is determined according to the following procedure.
The fraction of combinatoric events below the \Dsp peak is determined
from a fit to the \dm distribution in the region \mbox{$ 0.138 < \dm <
0.165~{\rm GeV}/c^2$} (see Figure \ref{f:dm}).  In the fit, \Dsp
events are described by the sum of two Gaussian functions while the
shape of combinatoric events is described by the function $\left[ 1 - \exp
\left( 
- \frac{\Delta m - m_{\pi} }{ c_1}
\right) 
\right] 
\left( 
\frac{\Delta m}{m_{\pi}}
\right)
^{c_2}$, where the overall normalization, $c_1$, and $c_2$ are
determined by the fit.  The width of the \dm distribution for signal
events is dominated by the experimental resolution.  In one third of
events the \pstar~ track is reconstructed in the SVT and in the DCH; in the
remainder, the track is reconstructed only in the SVT. As the
resolution in the first sample is better, the two samples are fitted
separately.  

The \dm\ distribution of a fake-lepton control sample is
fitted simultaneously to determine the fake-lepton background. The
control sample is obtained by selecting events with the kinematic cuts
described above, except for the requirement that the candidate lepton
fails a loose lepton selection criterion. Lepton identification
efficiencies and hadron misidentification probabilities, estimated
from data control samples of pure leptons and hadrons, are used to
scale the number of peaking events in this hadron-substituted sample to
the expected amount of fake-lepton background in the signal sample.

The continuum contribution is fitted from the off-resonance event
sample and scaled according to the ratio between the on-resonance and
off-resonance luminosities.  

For each \Dz\ final state, the signal,
fake-lepton and continuum samples are fitted simultaneously.  The mean
values, widths and the relative normalization of the two Gaussian
functions describing the signal are common to the three data sets,
while the parameters describing the shape of the combinatoric
background are fitted independently to each data sample.
Cuts on \dm are then applied to reduce the amount of combinatorics in
the subsequent stages of the analysis: 
\mbox{$0.143 < \dm < 0.148~{\rm GeV}/c^2$} if
the \pstar~ is reconstructed by the SVT alone, 
\mbox{$0.144 < \dm < 0.147~{\rm GeV}/c^2$} otherwise.

\begin{figure}[!t]
\begin{center}
\begin{tabular}{cc}
\includegraphics[width=0.49\textwidth, bb= 0 150 570 690, clip=]{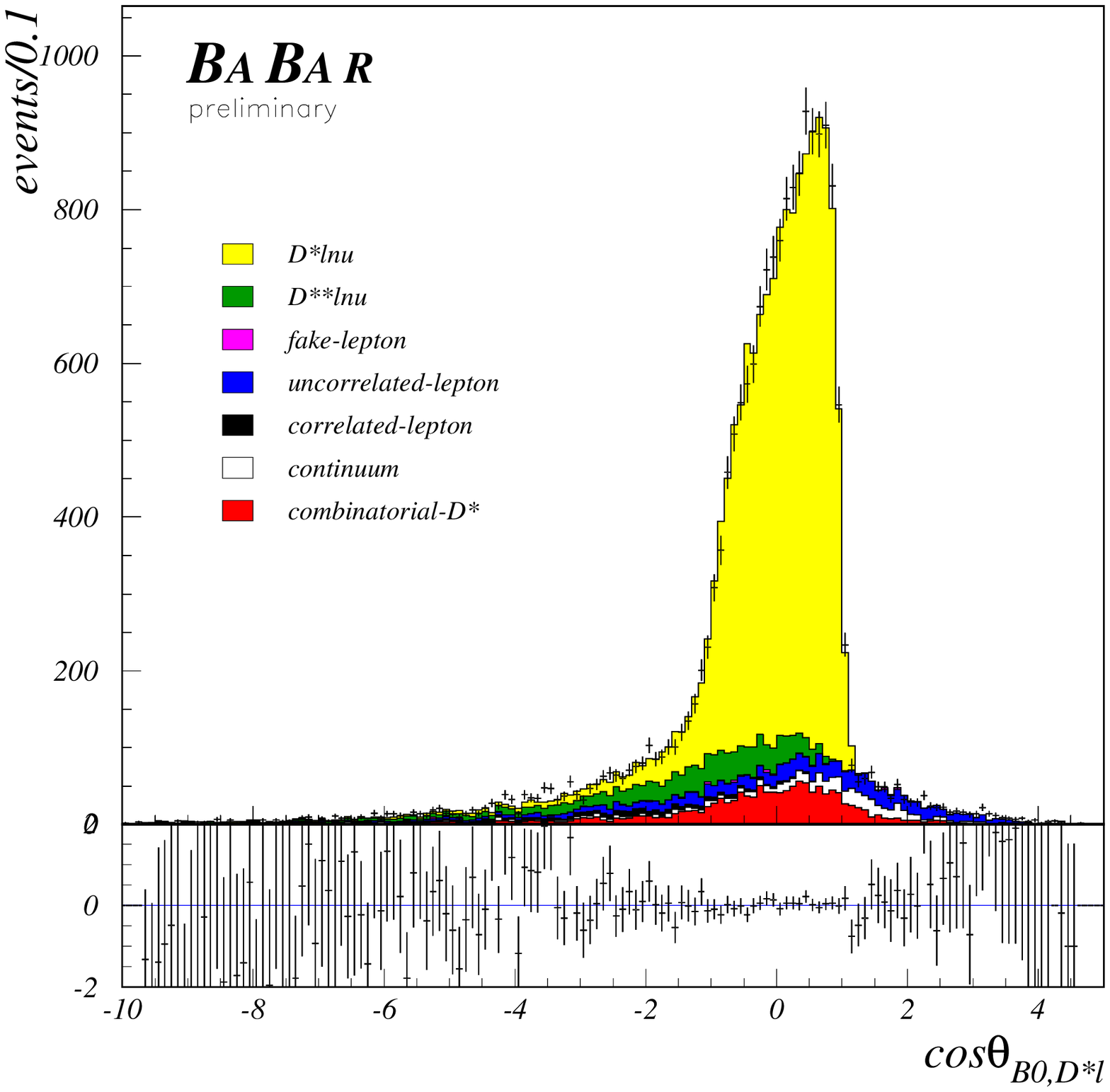} &
\includegraphics[width=0.49\textwidth, bb= 0 150 570 690, clip=]{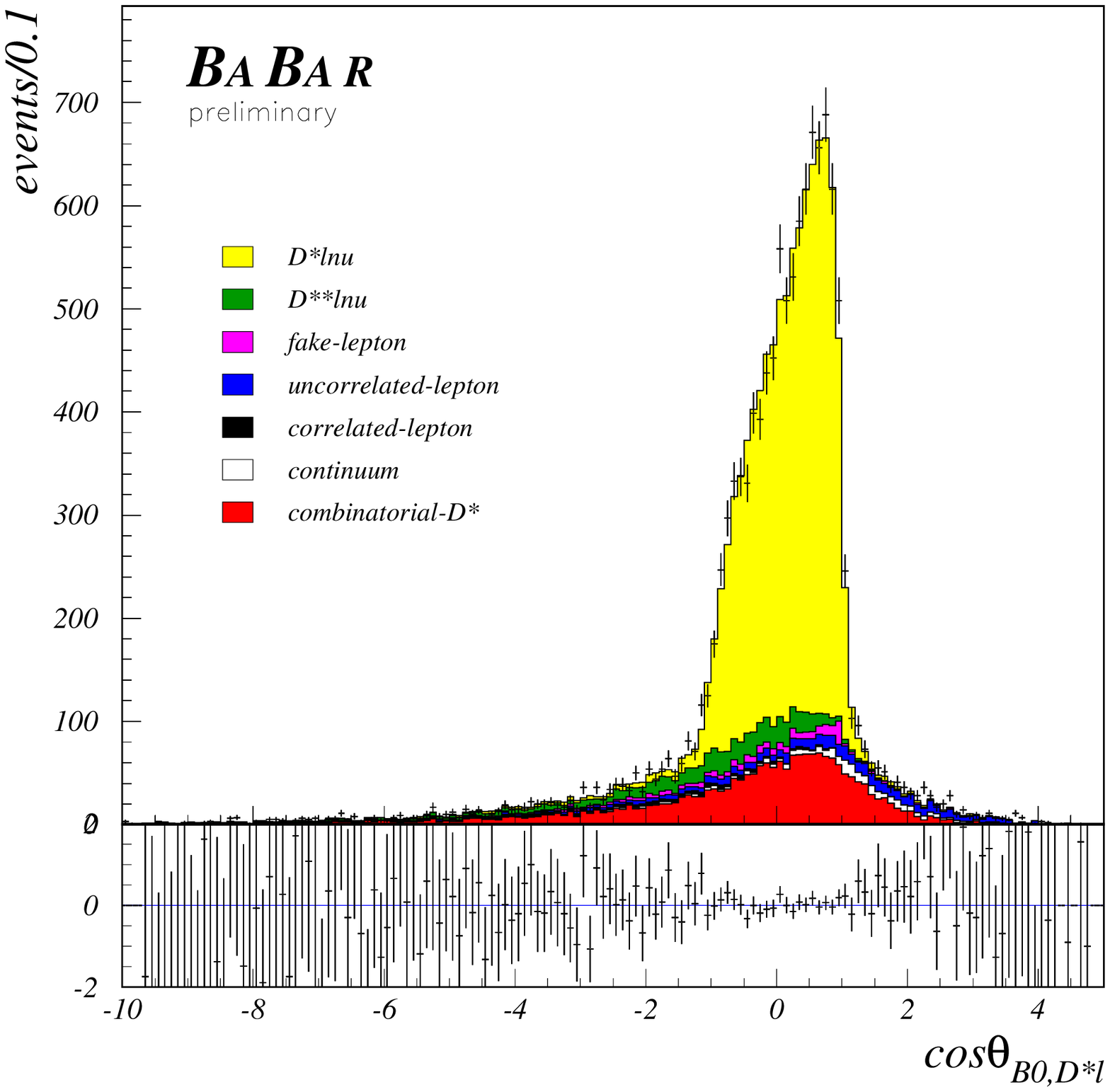} \\
\end{tabular}
\caption{Distributions in $\cos\TBY$ for the 
\mbox{$\bar{B^0} \rightarrow D^{*+} e^- \bar{\nu_e}, \Dz\ra\Kpi$} (left) and 
\mbox{$\bar{B^0} \rightarrow D^{*+} \mu^- \bar{\nu_\mu}, \Dz\ra\Kppz$} samples 
(right). Points with error bars show the data. Shaded areas represent
the Monte Carlo distributions. The fit determines the normalization of
the Monte Carlo distributions.  The tail to the left in the signal
electron sample is due to bremsstrahlung. Below the histograms are
shown the fractional deviations of the data from the fit results.}
\label{f:cosby}
\end{center}
\end{figure}

The fraction of uncorrelated and \Ddstar~ events is determined by
exploiting the decay kinematics.  The angle between the \Bz and the
pseudo-particle obtained by adding the \Dsp and the $\ell^-$
four-momenta is computed according to the following expression:
\ba
 \cos\TBY = \frac{ -(M^2_{\Bz} + M^2_{D^*\ell} -2E_{\Bz} E_{D^*\ell}) + M^2_{\rm miss} } { 2 p_{\Bz} p_{D^*\ell} } ,
\label{eq:ctby}
\ea
where $M_{\rm miss}$ is the invariant mass of all the other particles
produced in association with the lepton and the \Dsp in the decay of
the \Bz. We compute $\cos\TBY$ by fixing $M_{\rm miss}$ to zero, under the
hypothesis that this is a true \BtoDs\ decay, for which only a neutrino is
missing.  Neglecting the broadening due to experimental resolution,
the obvious constraint
\mbox{$-1 < \cos\TBY  < 1$} applies to signal events. Due to the production of additional 
particles, events from \Ddstar background have positive $M^2_{\rm
miss}$ values and therefore produce large tails below -1 in $\cos\TBY$
distribution, while uncorrelated background events preferentially
populate the region with $\cos\TBY>1$. This allows for the
determination of the fraction of \Ddstar, uncorrelated, and signal
events in the sample, by means of a fit to the $\cos\TBY$ distribution
in data (see for instance Figure \ref{f:cosby}).  The shapes of all
components in the fit are determined from Monte Carlo simulation,
while the number of events for each of the other background categories
is determined from the fit to the \dm\ distributions.  The same
procedure is applied to
%Data are separated into 18 independent samples, identified 
18 independent samples, identified 
according to lepton flavors ($\times 2$), period of data acquisition 
($\times 3$), \Dz decay mode ($\times 3$). 
%For the \Vcb~ determination, the data sets are further split into bins of \omt 
The data sets are further split into bins of \omt (defined in Section
\ref{sec:formfactor}) in order to reduce
the systematic error corresponding to variations in background
composition and normalization with $\omt$.  The resulting large
statistical uncertainties are reduced by requiring that the fractions
of uncorrelated and \Ddstar background with respect to the signal be
the same for electrons and muons, and for all \Dz~ decay modes. The
validity of these hypotheses is verified by Monte Carlo simulation.
The  distributions in $\cos\TBY$ for the
\mbox{$\bar{B^0} \rightarrow D^{*+} e^- \bar{\nu_e}, \Dz\ra\Kpi$} and 
\mbox{$\bar{B^0} \rightarrow D^{*+} \mu^- \bar{\nu_\mu}, \Dz\ra\Kppz$} samples 
(integrated over \omt) are shown in Figure~\ref{f:cosby}.

Only events with \mbox{$|\cos\TBY|  < 1.2$} are further analyzed.
This selects a final sample of 55700 signal events for the measurement.

\section{Determination of \boldmath{\Vcb}} 
\label{sec:fit}

\subsection{Parametrization of the Decay Width\label{sec:formfactor}}
Assuming the connection between the various form factors provided by HQET,
the expected number of signal events can be expressed as a function of
\om~ by the relation
\ba
\nonumber \frac{{\rm d}{\cal N}}{{\rm d}\om}~&=&~4~ N_{\Upsilon} ~ f_{00} ~ {\cal B}(\Dsp\ra \Dz\pstarp) ~ {\cal B}(\Dz\ra K^- n\pi) 
~\epsilon(\om)~  \frac{{\rm d}{\cal B}}{{\rm d}\om}, \\[3mm]
           \frac{{\rm d}{\cal B}}{{\rm d}\om}  ~ &=& ~ \dfrac{G^2_F}{48 \pi^3 ~ \tau_{\Bz}} M^3_{\Dsp} (M_{\Bz}-M_{\Dsp})^2 
                            ~\sqrt{\om^2-1}~ (\om+1)^2   \\
\nonumber  &\times&  \mid V_{cb} \mid^2 ~ {\cal F}^2(\om) 
                     \left[ 1+\frac{4\om}{1+\om}\frac{1-2\om r+r^2}{(1-r)^2}\right].
\ea
The factor of 4 accounts for the fact that two B mesons are produced in
each event, and that both electrons and muons are used; $N_{\Upsilon}$
is the number of $\Upsilon(4S)$ produced, $f_{00}$ is the fraction
$\Gamma(\Upsilon(4S) \ra \Bz\bar{\Bz}) /\Gamma(\Upsilon(4S))$
\footnote{The value $f_{00} = 0.487 \pm 0.013$ is obtained
from Table~\ref{tab:syst} assuming that the \FourS decays to only to
charged or neutral B mesons, i.e. imposing the constraint
$f_{00}+f_{+-} =1$.}, ${\cal B}(\Dsp\ra \Dz\pstarp)$ is the branching
fraction for the decay $\Dsp \ra D^0 \pi^+$, ${\cal B}(\Dz\ra K^-
n\pi)$ is the branching fraction for the decay of the \Dz to the final
state considered, $\tau_{\Bz}$ is the \Bz~ lifetime and $r$ is the ratio
of meson masses, $r = M_{\Dsp}/M_{\Bz}$. The values employed for these
parameters, as determined by independent measurements, are reported in
Table~\ref{tab:syst}.

\bt[hbt]
%\caption{\label{tab:syst} Values of the PDG normalization constants and their errors.}
\caption{\label{tab:syst} Values of inputs to the decay width calculation, 
as obtained from Ref.\cite{PDG}.}
\bc
\begin{tabular}{|l|l|}    
%\hline
%                &       PDG2002 value\\
\hline
Parameter & Value \\
\hline
${\cal B}(D^{*+}\to D^0\pi^+)$ &  $67.7 \pm 0.5  \%$  \\
%\hline
${\cal B}(D^0\to \Kpi)$        &  $3.80 \pm 0.09 \%$  \\
${\cal B}(D^0\to \Kppp)$       &  $7.46 \pm 0.31 \%$  \\
${\cal B}(D^0\to \Kppz)$       & $13.1  \pm 0.9  \%$  \\
$\tau_{\overline{B}_d}$        & $1.548 \pm 0.018$~ps \\
$f_{+-}/f_{00}$                & $1.055 \pm 0.055$    \\
%%$f_{00}$                & $0.487 \pm 0.013$    \\
\hline
\hline
\end{tabular}
\ec\et
The \om-dependent reconstruction efficiency $\epsilon(\om)$ is
determined by means of the detailed detector simulation.  The \om\
dependence of the form factor \Pfr(\om) is unknown.  Since only a  small
 range of \om\ is allowed by phase space, a Taylor series expansion limited
to second order has typically been used:

\be
{\cal F}(\om) ~=~ {\cal F}(1) ~ (1 + \rho^2_{\cal F} (1-\om) + c (1-\om)^2 + {\cal O}(1-\om)^3).
%\label{eq:Para1}
\eeq
%where the prediction that the first order coefficient must be positive~\cite{bho} 
%is made explicit by expressing it as a square real number. Contrary to 
Apart from \Pfr(1), the theory does not predict the values of the
higher order coefficients, which must be determined
experimentally. The first measurements of \Vcb~ were performed
assuming a linear expansion, i.e., setting $c=0$
%neglecting second order terms
(\cite{ARGUS,CLEO,ALEPH1,DELPHI:vcb}).  However, the
%Basic considerations derived from the 
requirement of analyticity and positivity of the QCD functions
describing the local currents leads to the prediction that $c$ should
% have a positive curvature which must be related to the radius of the
be positive and it should be related to the 
heavy meson radius \rhf (see~\cite{Neuold}) by the relation:
\be 
c ~=~ 0.66 {\rhf} - 0.11 ~.
\label{eq:Para2}
\eeq

Results employing this analyticity bound were obtained by the ALEPH,
DELPHI and OPAL collaborations (see~\cite{ALEPH2,DELPHI1,OPAL}). An
improved parametrization based on the above consideration is proposed
in~\cite{Neunew}, accounting for higher order terms and so reducing to
$\pm 2$\% the relative uncertainty on \Vcb~ due to the form factor
parametrization. A new function \Aone(\om) is introduced in that
calculation, connected to \Pfr(\om) by the following relation:
\ba
\label{eq:Para3}
 {\cal F}^2(\om)   ~&\times&~ \left[ 1+\frac{4\om}{1+\om}\frac{1-2\om r+r^2}{(1-r)^2}\right] ~=~ \\
\nonumber
 {\cal A}_1^2(\om) ~&\times&~ \left\{ 2\frac{1-2\om r+r^2}{(1-r)^2}
                                        \left[ 1+\frac{\om-1}{\om+1} R_1(\om)^2 \right] + 
                                        \left[ 1+\frac{\om-1}{1-r}(1-R_2(\om))\right]^2\right\} 
\ea 
where $R_1$(\om) and $R_2$(\om) are ratios of axial and vector form factors also given in Ref.~\cite{Neunew}. 
The following parametrization, depending only on a single unknown parameter \rha, 
is proposed for \Aone(\om):
\ba
\nonumber  {\cal A}_1(\om) ~=~ {\cal A}_1(1) \times \left[ 1-8\rho^2_{{\cal A}_1}z +(53\rho^2_{{\cal A}_1}-15)z^2-
      (231\rho^2_{{\cal A}_1}-91)z^3 \right]
\ea
with
\ba
\nonumber z = \frac{\sqrt{\om+1}-\sqrt{2}}{\sqrt{\om+1}+\sqrt{2}}.
\ea
It should be noted that, in the limit $\om\ra 1$, $\Aone(\om) \ra
\Pfr(\om)$, so that $\Aone(1)=\Pfr(1)$.\par Monte Carlo events
employed for this analysis were produced with a linear parametrization
for ${\cal F}(w)$, while experimental data are fitted using
expressions of Ref.~\cite{Neunew}.

%\subsection{Event Kinematics}
%\label{sec:kine}

Experimentally the variable \om\ can be expressed as:
\ba
\nonumber   \om = \frac{(M_{\Bz}^2 + M_{\Dsp}^2 - q^2)}{(2M_{\Bz} M_{\Dsp})}
\ea
 where $q^2 \equiv (p_{\Bz}-p_{\Dsp})^2$.
The magnitude of the \Bz momentum is known; its direction is obtained 
from Equation \ref{eq:ctby} with an azimuthal ambiguity about the
direction of the $\Dsp\ell^-$ pair. The two extreme solutions corresponding to the 
minimal and maximal angles between the \Bz and the $D^{*+}$ are used to define the quantity
\ba
\tilde{\om} \equiv \frac{\om_{\rm min} + \om_{\rm max}}{2}.
\ea 
The simulation shows that \omt is a good  estimator of \om, with a resolution 
\mbox{$\sigma(\tilde{\om}-\om) \sim 0.02$}, corresponding to about 4\% of the 
full physical range.

\subsection{Fit Method}
\label{fit-method}
Data and Monte Carlo simulated events are collected in ten equal-size \omt bins. 
A few events, which, due to resolution, have $\omt > \om_{\rm max} = 1.503$ are discarded. 
A least-squares fit is then performed comparing the number of events observed in each 
bin to the sum of signal and background events.
The number of background events in each bin is determined as explained above and is fixed 
in the fit. The signal contribution is obtained at each step in the minimization procedure 
by properly weighting each generated event surviving the selection. The $\chi^2$ is:
\ba
\label{eq:chi2}
\chi^2 = \sum_{i=1}^{10} \frac{(N^i_{\rm data} - N^i_{\rm bck}- \sum_{j=1}^{N^i_{\rm MC}} W_j )^2}
                                { N^i_{\rm data} + \sigma^2_{\rm bck} +  \sum_{j=1}^{N^i_{\rm MC}} W_j^2},
\ea
where the index $i$ runs over the ten \omt bins; $N^i_{\rm data (MC)}$ is the number of data 
(signal Monte Carlo)
events found in the $i^{th}$ bin; $N^i_{\rm bck}$, $\sigma^i_{\rm bck}$ are the numbers of 
background events and their errors. The weight for each Monte Carlo event is computed 
as the product of three terms
\ba
W_j = W^{\cal L} \times W^{\epsilon}_j \times W^{theo}_j,
\ea
where

\bi

\item $W^{\cal L}$ is an overall fixed scale factor, that accounts for the relative luminosity of 
data and signal Monte Carlo events, for the difference in the actual branching fractions 
relevant to this analysis~\footnote{obviously not including ${\cal B }(\BtoDs)$}, and any possible 
overall efficiency scale factor.

\item $ W^{\epsilon}_i$ accounts for the efficiency correction, computed event per event as the ratio 
of the efficiency in the real data and in the Monte Carlo. 
It accounts for differences in tracking, particle identification, $\pi^0$ reconstruction, 
and depends on the three-momenta of the lepton, the soft pion and of all the particles 
from the \Dz decay. It remains unchanged at all steps of the minimization. 
(If the Monte Carlo simulation were perfectly tuned, this factor would be identically one). 
It should be noted that only the factor due to the slow pion tracking efficiency introduces 
a net dependence of this term on $\omt$.

\item $ W_i^{theo} = f_{\rm theo}(\om ; \rho^2, \Vcb ) / f_{\rm MC}(\om ; \rho^2_{\rm MC}, \Vcb_{\rm MC} ) $
is the term accounting for the theoretical function describing the decay. It depends on the 
parameters to be determined (\AoneVcb\ and \rha ) and varies for each step of the minimization.
\ei
Here the function $f_{\rm theo}$ corresponds to the expressions in Equations 3 
and~\ref{eq:Para3}, while the function used to generate
Monte Carlo events, $f_{MC}$, contains a simpler, linear parametrization 
of the form factor dependence on $\om$. 

The value of the branching ratio is then computed by integrating the differential expression in Equation 3. 

The fit has been performed as a blind analysis, i.e., the values of \AoneVcb\ and \rha\ were hidden until the 
study of the systematic errors was completed and all consistency checks were performed.
%Bob

\subsection{Fit Results}

\begin{figure}[!ht]
\begin{center}
\includegraphics[width=3.2in,height=3.3in]{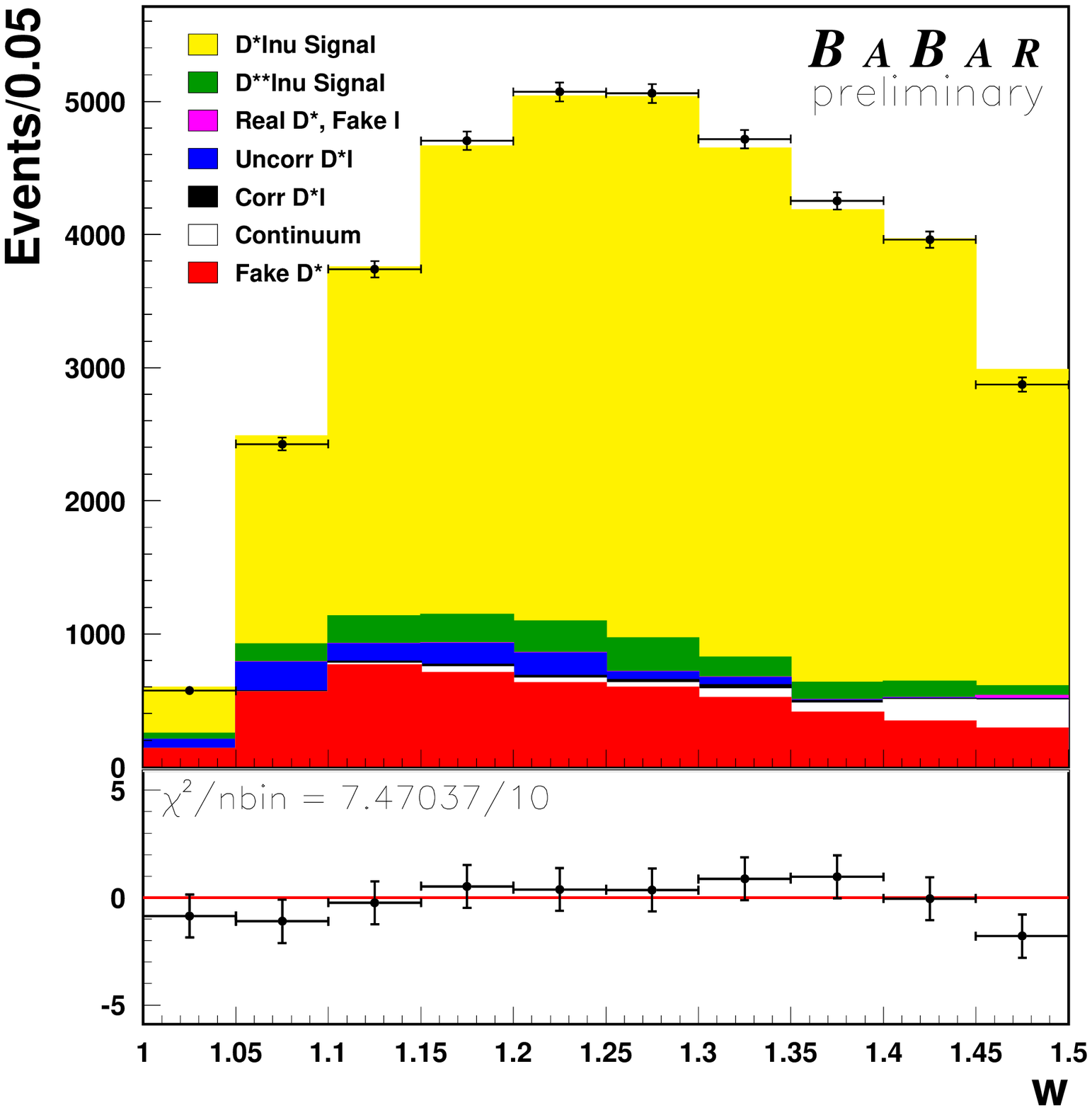}
\includegraphics[width=3.2in,height=3.3in]{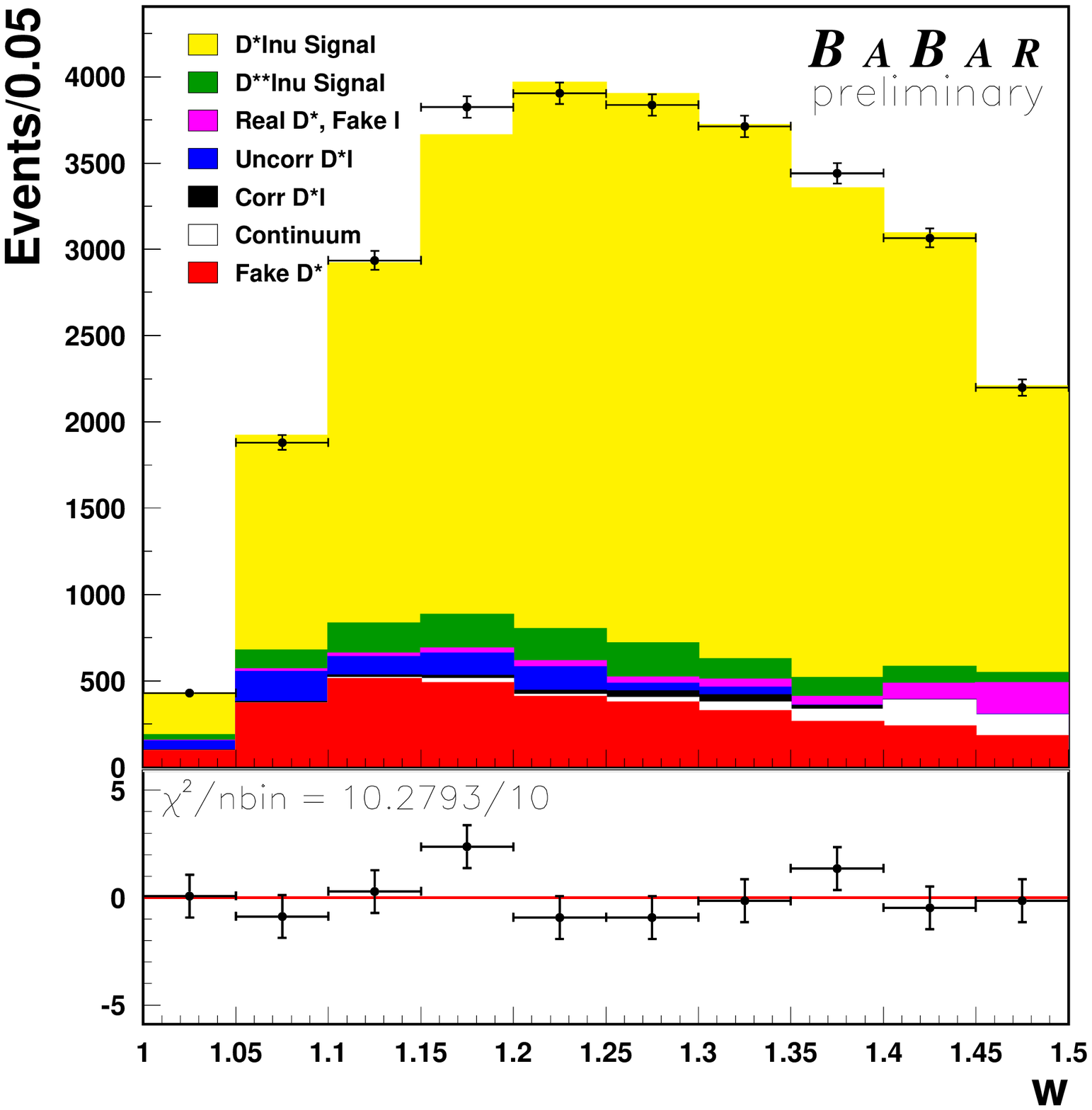}\\
\caption{ The experimental distribution of \om\ compared to fit results separately for electrons (left) and muons (right). 
The plots from all the different $D^0$ decay modes and years of data taking are summed together. 
The points are the data and the histograms are the results of the fit to the sum of the signal and the different sources of background. 
The fit residuals are also shown below.}
\label{f:wreco_combo}
\end{center}
\end{figure}

The fit is performed separately for each data set, providing 18 statistically independent 
determinations of \AoneVcb\ and of \rha. The results are reported in Table~\ref{t:results}.
%together with their statistical correlation.
In the same Table, the branching fraction, as obtained by integrating the differential
decay width, is also shown. The agreement between the data and
the fit function is good. The distribution of the $\chi^2$ probabilities is uniform. 
The measured \om\ distributions are shown in Figure~\ref{f:wreco_combo}, where the 
three distributions for the \Dz decay modes are summed.

It should be noted that, because the statistical errors are small, the measurements 
are dominated by the systematic errors, which introduce considerable correlations among the samples. 
For this reason, first the systematic error is discussed, and then the average is presented.

%\begin{figure}[!t]
%\begin{center}
%\begin{tabular}{ c }
%\includegraphics[width=7.2cm]{probchi2.eps}\\
%\end{tabular}
%\caption{\label{f:prob} Distribution of the $\chi^2$ probability for the different data sets. The mean and RMS values of this distribution are compatible with the expected values of an uniform distribution.}
%\end{center}\end{figure}

%----- results 2000 2001 2002 ---------------------------------------------------------------------

\bt[!t]\bc
\caption{\label{t:results} Fit results and the branching ratio \mbox{${\cal B}(\BtoDs)$} , for 
  sub-samples defined by $D^0$ decay mode, lepton species, and year of data taking. The third columns
shows the statistical correlation between \AoneVcb\ and  \rha. }
\vs{0.3cm}\begin{tabular}{|c|c|cccc|c|}
\hline  \hline
% $D^0$ mode        & $\AoneVcb\times 10^3$ &  \rha & $C_{\AoneVcb,\rho^2}$ $\%$ & {\cal BR} $\%$ & P($\chi^2$) $\%$ \\
 & $D^0$/lepton        & $\AoneVcb\times 10^3$ &  \rha & Corr$_{stat}$ & ${\cal B}$ $\%$ & P($\chi^2$) $\%$ \\
\hline\hline 
 & $K\pi$/e            &  34.69$\pm$0.95 &   1.284$\pm$ 0.083 &   91$\%$ &   4.72$\pm$ 0.11 &   \phantom{0}3.1 \\ 
 & $K\pi\pi\pi$/e      &  35.68$\pm$1.42 &   1.452$\pm$ 0.106 &   93$\%$ &   4.55$\pm$ 0.16 &   98.4 \\ 
\rotatebox{90}{\hspace{-5mm}2000}
 & $K\pi\pi^0$/e       &  32.36$\pm$1.17 &   1.241$\pm$ 0.109 &   93$\%$ &   4.21$\pm$ 0.12 &   20.7 \\ 
\cline{2-7}
 & $K\pi$/$\mu$        &  33.00$\pm$0.99 &   1.145$\pm$ 0.097 &   92$\%$ &   4.61$\pm$ 0.11 &   21.6 \\ 
 & $K\pi\pi\pi$/$\mu$  &  33.51$\pm$1.48 &   1.410$\pm$ 0.127 &   93$\%$ &   4.11$\pm$ 0.15 &   26.0 \\ 
 & $K\pi\pi^0$/$\mu$   &  32.75$\pm$1.20 &   1.235$\pm$ 0.111 &   93$\%$ &   4.32$\pm$ 0.13 &   83.1 \\ 
\hline 
\hline
 & $K\pi$/e            &  34.92$\pm$0.73 &   1.237$\pm$ 0.063 &   92$\%$ &   4.91$\pm$ 0.08 &   \phantom{0}6.8 \\ 
 & $K\pi\pi\pi$/e      &  35.06$\pm$1.08 &   1.357$\pm$ 0.085 &   93$\%$ &   4.63$\pm$ 0.12 &   45.6 \\ 
\rotatebox{90}{\hspace{-5mm}2001}
 & $K\pi\pi^0$/e       &  33.68$\pm$0.89 &   1.190$\pm$ 0.081 &   93$\%$ &   4.69$\pm$ 0.09 &   70.1 \\ 
\cline{2-7}
 & $K\pi$/$\mu$        &  34.35$\pm$0.78 &   1.242$\pm$ 0.070 &   92$\%$ &   4.74$\pm$ 0.09 &   25.6 \\
 & $K\pi\pi\pi$/$\mu$  &  34.83$\pm$1.16 &   1.407$\pm$ 0.092 &   93$\%$ &   4.45$\pm$ 0.13 &   41.8 \\ 
 & $K\pi\pi^0$/$\mu$   &  33.69$\pm$0.96 &   1.191$\pm$ 0.088 &   93$\%$ &   4.69$\pm$ 0.10 &   72.0 \\ 
\hline 
\hline
 & $K\pi$/e            &  33.75$\pm$0.89 &   1.212$\pm$ 0.082 &   92$\%$ &   4.65$\pm$ 0.10 &   68.8 \\ 
 & $K\pi\pi\pi$/e      &  33.89$\pm$1.32 &   1.409$\pm$ 0.105 &   93$\%$ &   4.21$\pm$ 0.14 &   73.2 \\ 
\rotatebox{90}{\hspace{-5mm}2002}
 & $K\pi\pi^0$/e       &  35.41$\pm$1.06 &   1.364$\pm$ 0.085 &   92$\%$ &   4.71$\pm$ 0.12 &   93.3 \\ 
\cline{2-7}
 & $K\pi$/$\mu$        &  32.71$\pm$1.06 &   1.034$\pm$ 0.107 &   93$\%$ &   4.81$\pm$ 0.12 &   37.6 \\ 
 & $K\pi\pi\pi$/$\mu$  &  33.84$\pm$1.52 &   1.323$\pm$ 0.129 &   93$\%$ &   4.40$\pm$ 0.17 &   45.6 \\
 & $K\pi\pi^0$/$\mu$   &  35.52$\pm$1.24 &   1.406$\pm$ 0.101 &   91$\%$ &   4.63$\pm$ 0.14 &   42.9 \\ 
\hline \hline
\end{tabular}

\ec\et

\subsection{Systematic Errors and Consistency Checks}
\label{cha:sys}
The individual sources of systematic errors are reported in
Table~\ref{t:syste} and are described in detail below. They can be
grouped as
\bi
%\item Factors affecting the overall normalization (see Equation \ref{eq:width}): 
\item Factors affecting the overall normalization (see Equation 3): 
Each of these factors is varied by its uncertainty, and the resulting variation in the
parameters is added to the systematic error.  The uncertainties on
$N_{\Upsilon}$, $f_{00}$, $\tau_{\Bz}$ and ${\cal B}(\Dsp)$ are
completely correlated for all the samples. The uncertainties on the
$\Dz$ branching fractions depend on the decay mode and are only
partially correlated~\cite{PDG}. % The correlation coefficients are taken from
% the PDG~\cite{PDG}.
\item Experimental uncertainties in the
  efficiency: These include track reconstruction, lepton and
  kaon identification, $\pi^0$ selection, vertex reconstruction and
  the cut on the Dalitz plot for the \Kppz mode. \\ For high-momentum
  tracks, the tracking efficiency is determined
  comparing the independent information from the SVT and 
  DCH. This approach results in $\pm 0.8\%$ systematic uncertainty in the
 reconstruction efficiency  for each
  track; the errors from each track are added linearly (i.e., the
  error on the branching fraction is $N$ times the single track error,
  where $N$ is the total number of tracks employed, including the
  lepton). Most of the \pstar\ do not reach the DCH, and are
  reconstructed in the SVT alone. The efficiency for these
  low-momentum tracks is computed from the angular distribution of
  the \pstar\ in the \Dsp rest frame. To evaluate this, a large set of
  $\Dsp \ra \Dz \pstarp$, $\Dz \ra \Kpi$ decays is selected from
  generic hadronic events. For fixed values of the \Dsp\ momentum, the
  observed angular distribution is compared to the one predicted for
  the decay of a vector meson to two pseudoscalar mesons, and the
  deviation from the expected shape is attributed to the inefficiency
  of track reconstruction. This inefficiency is parametrized as a
  function of the particle momentum in the laboratory frame.
%  The efficiency of low-momentum  tracks is therefore
%  related to the one of the stiffer tracks  in the sample. These last
%  are hard enough that their absolute efficiency is known (see
%  above). 
  The values of the parameters describing the efficiency
  are computed in different bins of the  polar angle of the
  track, both in the data and in the simulation.  The systematic error
  is computed by varying each parameter of the
  efficiency function by its uncertainty, including correlations.

The efficiencies and systematic errors for lepton and kaon identification and $\pi^0$
reconstruction are measured from control samples. The error on lepton
identification is common to all samples (separately for each lepton type). 
Looser kaon identification criteria are imposed on \Kpi decays; the 
corresponding systematic error is therefore smaller than for \Kppp and \Kppz decays. \\
The systematic errors due to the cuts on the vertex probability and on 
the Dalitz plot are evaluated by varying these cuts.\\
The efficiency corrections in data and in Monte Carlo
simulation are largely independent of $\omt$, and therefore the value of \rha\ is scarcely affected, while a substantial error is induced on \AoneVcb~ and branching fraction. 
Not surprisingly the most noticeable exception is the \pstar\ 
tracking efficiency correction, which affects both \AoneVcb\ and \rha.

\begin{table}[!ht]
\caption{Decay modes used in the Monte Carlo simulation of the
$\bar{B}\ra\Dstarp\ell^-\bar{\nu_\ell} X $.
These branching fraction values are based mainly on theoretical estimates.
The theoretical models adopted to generate these events are also listed.
For the four-body $\bar{B}\ra\Dstarp\ell^-\bar{\nu_\ell}\pi$ decays, the Goity and
Roberts model (GR)~\cite{ref:goity} was used, while the Isgur and Scora 
model (ISGW2)~\cite{ref:IGSW2} was adopted for the resonant
$\bar{B}\ra\Ddstar\ellp\nul$ decays.} 
\begin{center}
\begin{tabular}{|c|c|c|c|c|c|}
\hline
\B decay mode                                               &     ${\cal B}(B)$        &  Model&  \Ddstar decay      & ${\cal B}(D)$ & Overall BR  \\ 
                                                    & $\times10^{-2}$ &         &  &    & $\times10^{-4}$\\
\hline
\hline
$\Bz \rightarrow D^{* -}\pi^0 \ellp \nul $   &   0.10          & GR      &        $-$                                 & $-$   &   10.           \\ 
$\Bz \rightarrow D^-_1 \ellp\nul             $   &   0.56          & ISGW2   &$D^-_1\rightarrow D^{* -}\pi^0$        & 0.33 &  18.5          \\ 
$\Bz \rightarrow D^{* -}_1 \ellp\nul     $   &   0.37          & ISGW2   &$D^{* -}_1\rightarrow D^{* -}\pi^0$& 0.33 &  12.2          \\ 
$\Bz \rightarrow D^{* -}_2 \ellp\nul     $   &   0.37          & ISGW2   &$D^{* -}_2\rightarrow D^{* -}\pi^0$& 0.103&  3.81           \\ 
$\Bz \rightarrow D^{\prime -} \ellp\nul      $   &   0.02          & ISGW2   &$D^{\prime -}\rightarrow D^{* -}\pi^0$ & 0.33 &  0.67           \\ 
$\Bz \rightarrow D^{* \prime -} \ellp\nul$   &   0.22          & ISGW2   &  $D^{* \prime -}\ra\Dstarm\piz$       & 0.17   & 3.74                \\ 
\hline
$\Bp \rightarrow D^{* -}\pip \ellp\nul $         &   0.20          & GR      &        $-$                                 & $-$   &    20.          \\ 
$\Bp \rightarrow D^0_1 \ellp\nul             $         &   0.56          & ISGW2   &$D^0_1\rightarrow D^{* -}\pip$        & 0.67 &   37.5         \\ 
$\Bp \rightarrow D^{* 0}_1 \ellp\nul     $         &   0.37          & ISGW2   &$D^{* 0}_1\rightarrow D^{* -}\pip$& 0.67 &   24.8        \\ 
$\Bp \rightarrow D^{* 0}_2 \ellp\nul     $         &   0.37          & ISGW2   &$D^{* 0}_2\rightarrow D^{* -}\pip$& 0.21 &   7.78          \\ 
$\Bp \rightarrow D^{\prime 0} \ellp\nul      $         &   0.02          & ISGW2   &$D^{\prime 0}\rightarrow D^{* -}\pip$ & 0.67 &   1.32          \\ 
$\Bp \rightarrow D^{* \prime 0} \ellp\nul$         &   0.22          & ISGW2   & $D^{* \prime 0}\ra\Dstarm\pip$                          & 0.33    &  7.26
               \\ 
\hline
\end{tabular}
\end{center}
\label{t:dss}
\end{table}

\item Background subtraction: The fraction of background events is determined 
in data for each \omt bin as described above.  This procedure
considerably reduces the systematic error.  The statistical error on
the background is automatically accounted for by the fit (see
Equation~\ref{eq:chi2}).  A residual model dependence is considered
for the \Ddstar background.  Including narrow, wide and non-resonant
$\Dsp\pi$ states, a total of twelve modes (for \Bz and \Bu
semileptonic decays) is considered. The \Ddstar background is computed
assuming the branching fractions reported in Table~\ref{t:dss}.  In
order to estimate the systematic error due to the
uncertainty in the shape of the \Ddstar background, the $\cos\TBY$ fits
in the different \om\ bins are performed considering only one decay
mode a time, and the corresponding values of \Vcb and of \rha are then
determined again. The systematic error is then computed as half the
difference between the maximum and minimum values of the parameters
obtained from this study.\\ To test whether the background from
uncorrelated events is properly handled, the cut on
$\cos\theta_{D^*\ell}<0$ is removed and the study is repeated. This
results in an increase of the uncorrelated background by a factor of
about 2 in the full $\cos\TBY$ range, and by 30\% in the signal
region. No appreciable change in the fit result is observed.  As an
alternative, the fraction of uncorrelated events is determined by
counting the events in the rejected region $\cos\theta_{D^*\ell}>0$
and propagating this number to the complementary region using the
Monte Carlo. The fraction obtained is then fixed in the $\cos\TBY$
fit, which is used to determine the fraction of \Ddstar and signal
events only. Once again, no appreciable change in the result is
observed.\\ Finally, the hypothesis that the ratio of the amount of
\Ddstar and uncorrelated background over the signal be the same for
all \Dsp decay modes is checked with the simulation.  While the test
is satisfactory for the uncorrelated background, a slight discrepancy
is observed for \Ddstar events. For \Kppp decay the fitted fraction is
higher by 30\% than for the other modes. The effect is the same
for electron and muon events and does not depend on $\omt$.  Therefore
the \omt fit is repeated by increasing the \Ddstar amount in the \Kppp
sample and decreasing it correspondingly in the \Kpi and \Kppz. The
difference in the result is propagated as systematic error.

\item Theoretical uncertainties: A considerable fraction of the allowed phase-space is removed 
by the lower limit on the lepton momentum. This induces 
a model-dependence in the computation of the branching fraction. The lepton spectrum is 
determined partly by the fitted shape of the Isgur-Wise function, and partly 
by the angular profile of the decay, which is governed by the parameters $R_1$ and $R_2$
defined above. Even with perfect acceptance, the  uncertainty on $R_1$ and $R_2$ directly 
affects \rha\ and \AoneVcb, because these parameters enter in the fit function. 
The values of these 
parameters determined by the CLEO collaboration~\cite{ref:CLEOff} are used 
in this analysis. The systematic error is taken as the observed variation of \AoneVcb, \rha\ 
and of ${\cal B }(\BtoDs)$, when $R_1$ and $R_2$ are floated by their uncertainties, accounting for 
their correlation. 

\item Fit method: The fit procedure is validated using an independent sample of Monte Carlo
simulated events, of the same size as the data, and following the
procedure described above, using however a linear extrapolation for
the Isgur-Wise function.  The fitted values of \rha\ and \AoneVcb\ are
consistent within their statistical errors with the input values.  The
statistical error on this test is added to the systematic error.  In
addition, to test that the fit method does not bias the result, a set
of 500 toy experiments are performed, in which random events are
generated with realistic efficiency and resolution.  Each toy
experiment has the same sample size for data and Monte Carlo as in the
actual measurement. Within errors, no difference is observed between
the average of the fitted parameters and their generated values. Also,
it is observed that the pull width is consistent with one.
\ei

\begin{table}[htb]
\caption{Summary of all statistical and systematic errors.}
\begin{center}
\begin{tabular}{|l|c|c|c|}
\hline\hline
%error contribution & $\Delta \AoneVcb \times 10^3$ ($\%$)   & $\Delta$\rha  & $\delta {\cal B} / {\cal B}$ ($\%$) \\
error contribution & $\delta\AoneVcb / \AoneVcb$ ($\%$)   & $\delta$\rha  & $\delta {\cal B} / {\cal B}$ ($\%$) \\
\hline\hline
statistics (data and MC)                & {\bf 0.7}   & {\bf 0.02}  & {\bf 0.8} \\
\hline\hline
particle identification                 & 0.5   & -     & 0.9 \\
\pstar\ efficiency                      & 1.3   & 0.02  & 1.9 \\
tracking $\&$ $\pi^0$ efficiency        & 1.3   & -     & 2.7 \\

$\Dstarp\ell^-$ vertexing efficiency    & 0.5   & -     & 1.0 \\
\om~ fit method				& 0.6	& 0.02	& 1.2 \\
$\cos\TBY$ \chisq fit binning           & 0.5   & -     & 1.0 \\
$D^{**}$ \bkg\ composition              & 1.8   & 0.06  & 2.0 \\
total number of $B$ produced            & 0.6   & -     & 1.1 \\
\FourS rest frame \B momentum           & 0.3   & -     & 0.7 \\
$R_1(1)$ and $R_2(1)$                   & 1.8   & 0.27  & 1.8 \\
\hline\hline
total systematic error                  &{\bf 3.4}& {\bf 0.28}& {\bf 5.0} \\
\hline\hline
$\tau_{\Bz}$				& 0.6	& -	& -   \\
${\cal B}(\Dz)$                         & 1.1   & -     & 2.0 \\
${\cal B}(\Dstarp\ra\Dz\pip)$           & 0.4   & -     & 0.7 \\
${\cal B}(\FourS\ra\BzBzb)$             & 1.3   & -     & 2.7 \\
\hline\hline
total systematic error                  &{\bf 1.8}& - & {\bf 3.5} \\
\hline\hline
\end{tabular}
\end{center}
\label{t:syste}
\end{table}

\subsection{Results}

The results from all subsamples are combined by using the {\sc COMBOS}
package~\cite{combos}, taking into account the correlations between
the samples. {\sc COMBOS} was originally developed by the
LEP-B-Oscillation working group to combine the results of $\Delta M_d$
from different experiments, and was then adapted by the LEP-V$_{cb}$
working group to compute the world average for $|V_{cb}|$, \rha\ and
${\cal B}(\BtoDs)$. Besides computing averages, errors, and confidence
levels, {\sc COMBOS} also provides a breakdown of the single error
sources, which is reported in Table~\ref{t:syste}.

All the parameters affecting the normalization are treated as fully
correlated.  The errors on the branching fractions of the \Dz to each
final state are fully correlated for each state, but, at present, the
correlations among different decay modes are neglected.  The systematic
errors on particle identification are considered as fully correlated
among data taking periods.  The error on tracking is determined by
systematic effects common to all years and therefore is considered as
fully correlated among all the samples. In contrast the error on
\pstar\ tracking efficiency is still dominated by the statistical
uncertainty from the control samples; therefore it is completely
correlated for samples collected in the same year of data acquisition,
but uncorrelated for different years. Model errors are common to all
data sets.\\ The mean values are obtained by first averaging results
for different running periods, and then computing the average result
for the three years.  The confidence level of the \AoneVcb\ and \rha\
average values is 30\%.  The results are reported in
Table~\ref{t:average-years}.

\bt[!ht]
\caption{\label{t:average-years} Fit results for the three running periods.}
\bc
\begin{tabular}{|l|cc|}
\hline\hline
year            &$\AoneVcb\times 10^3$           & $\rho^2_{{\cal A}_1}$  \\
\hline 
2000            &33.60$\pm$0.46$\pm$1.33     &1.23$\pm$0.04$\pm$0.28           \\
2001            &34.63$\pm$0.36$\pm$1.44     &1.25$\pm$0.03$\pm$0.28           \\
2002            &34.73$\pm$0.45$\pm$1.39     &1.31$\pm$0.04$\pm$0.28           \\
\hline
Total Average   &34.03$\pm$0.24$\pm$1.31     &1.23$\pm$0.02$\pm$0.28           \\
\hline\hline
\end{tabular}
\ec\et

\noindent
For the sum of all 18 data subsamples the following result is obtained:
\ba 
\label{eq:results}
\AoneVcb  &=& (34.03 \pm 0.24 \pm 1.31)\times 10^{-3}, \\
\rha             &=&   1.23 \pm 0.02 \pm 0.28, \\ 
{\cal B}(\BtoDs) &=&  (4.68 \pm 0.03 \pm 0.29)~\%,
\ea
where the first error is statistical and the second is systematic. The statistical correlation 
between \AoneVcb and \rha is 92$\%$.

%To control the internal consistency of the result, the average has 
%been performed by grouping samples by lepton kind or by \Dz\ decay mode.
%Always a good consistency is found (see tables \ref{t:average-lepton,t:average-mode}).
%It should be noted that if the average is computed by grouping events 
%first by decay mode, a small difference is observed compared to what reported in
%\ref{eq:results}. The difference, corresponding to a fraction of the overall error,
%is attributed to numerical effects and is added to the systematic error.

\section{Conclusions \label{conc}}
A sample of 55,700 signal events from the decay process
\BtoDs\ is selected from a set of $86 \times 10^6$ \upsbb decays
collected by the \babar\ detector.
%The slope and the normalization of the form factor of the decay
%are measured from a fit to the differential branching fraction
%\mbox{${\rm d}{\cal B}/{\rm d}\om$}. 
The product of \Vcb\ and the form factor at zero recoil,
${\cal F}(1) = {\cal A}_1(1)$, is measured, as well as the derivative of the form factor,
\rha , again at zero recoil.
The integrated branching fraction is 
also computed. 
These results are in agreement with those obtained 
by Belle~\cite{BELLE} and with the LEP average~\cite{PDG}, but 
differ by more than two standard deviations from the CLEO result~\cite{CLEO}.
Using the value ${\cal A}(1) = 0.913^{+0.030}_{-0.035}$ reported
in~\cite{Hashimoto}, based on a lattice QCD calculation, the value:
\begin{eqnarray} 
 \Vcb &=& (37.27\pm0.26 \mathrm{(stat.)} \pm1.43 \mathrm{(syst.)} ^{+1.48} _{-1.23} \mathrm{(th.)} ) \times 10^{-3}
\end{eqnarray}
is then computed. 

From the measurements of the hadronic mass moments and decay rate for the inclusive 
semileptonic $B$ decays \babar\ has obtained another determination of 
$\Vcb$~\cite{moments}. The preliminary result is:
\begin{eqnarray} 
 \Vcb &=& (42.1\pm1.04 \mathrm{(exp.)} \pm0.72  \mathrm{(th.)} ) \times 10^{-3}
\end{eqnarray}
Both the experimental and theoretical errors are independent. By interpreting
the theoretical errors as $\pm 68\%$ confidence interval, the two
values differ by 2.0 standard deviations.

% Standard acknowledgments paragraph; must always be included.
\section{Acknowledgments}

We are grateful for the 
extraordinary contributions of our \pep2\ colleagues in
achieving the excellent luminosity and machine conditions
that have made this work possible.
The success of this project also relies critically on the 
expertise and dedication of the computing organizations that 
support \babar.
The collaborating institutions wish to thank 
SLAC for its support and the kind hospitality extended to them. 
This work is supported by the
US Department of Energy
and National Science Foundation, the
Natural Sciences and Engineering Research Council (Canada),
Institute of High Energy Physics (China), the
Commissariat \`a l'Energie Atomique and
Institut National de Physique Nucl\'eaire et de Physique des Particules
(France), the
Bundesministerium f\"ur Bildung und Forschung and
Deutsche Forschungsgemeinschaft
(Germany), the
Istituto Nazionale di Fisica Nucleare (Italy),
the Foundation for Fundamental Research on Matter (The Netherlands),
the Research Council of Norway, the
Ministry of Science and Technology of the Russian Federation, and the
Particle Physics and Astronomy Research Council (United Kingdom). 
Individuals have received support from 
the A. P. Sloan Foundation, 
the Research Corporation,
and the Alexander von Humboldt Foundation.

\end{document}